\documentclass[a4paper,11pt]{article}

\usepackage{fullpage}
\usepackage{times}
\usepackage{soul}
\usepackage{url}
\usepackage[utf8]{inputenc}
\usepackage[small]{caption}
\usepackage{graphicx}
\usepackage{xcolor}
\usepackage{amsmath}
\usepackage{booktabs}
\usepackage{algorithm}
\usepackage{enumitem}

\usepackage{amsthm}
\usepackage{amssymb}
\usepackage{algpseudocode}
\usepackage{cleveref}
\usepackage{mathtools}

\usepackage{bbm}

\newtheorem{theorem}{Theorem}
\newtheorem{corollary}[theorem]{Corollary}

\newtheorem{lemma}[theorem]{Lemma}

\theoremstyle{definition}

\usepackage{natbib}

\newcommand{\owner}{\mathbf{owner}}
\newcommand{\held}{\mathbf{held}}
\newcommand{\util}{\mathbf{util}}
\newcommand{\pairs}{\mathbf{pairs}}
\newcommand{\h}{\mathbf{h}}
\newcommand{\size}{\mathbf{size}}
\newcommand{\pos}{\mathbf{pos}}

\allowdisplaybreaks

\usepackage{authblk}

\title{\bf On the Complexity of Envy Cycle Elimination}

\author{Warut Suksompong}

\affil{National University of Singapore}

\date{\vspace{-10mm}}

\begin{document}

\maketitle

\begin{abstract}
The envy cycle elimination (ECE) algorithm is a fundamental algorithm for the fair allocation of indivisible goods, guaranteeing an allocation that is envy-free up to one good under arbitrary monotone utilities.
We show that the ECE algorithm can be implemented in $O(nm)$ time for $n$~agents and $m$~goods, and establish the tightness of this bound.
\end{abstract}

\section{Introduction} 

The fair allocation of resources is an important problem in everyday life and has been studied by researchers across several disciplines \citep{BramsTa96,Moulin03}.
A ubiquitous application is when the resources consist of indivisible goods, such as houses, cars, or paintings, where each good needs to be allocated as a whole \citep{BouveretChMa16,AmanatidisAzBi23}.

While numerous approaches for allocating indivisible goods fairly have been proposed in the literature, one of the most fundamental is the \emph{envy cycle elimination (ECE)} algorithm \citep{LiptonMaMo04}.\footnote{The work of \citet{LiptonMaMo04} has been cited almost 1,000 times according to Google Scholar and received the ACM SIGecom Test of Time Award in 2025.}
The ECE algorithm operates by arranging the goods in arbitrary order and allocating each good to an unenvied agent.
In order to ensure that some agent is unenvied and can receive the next good, any envy cycle that arises is eliminated by giving the bundle of each agent in the cycle to the agent who envies her.
The algorithm always produces an allocation that satisfies \emph{envy-freeness up to one good (EF1)}: this means that if an agent envies another agent, then the envy can be resolved by removing some good from the latter agent's bundle.
In fact, this guarantee holds even when agents have arbitrary monotone utilities.
This stands in contrast to other well-known approaches such as maximum Nash welfare or round-robin, which ensure EF1 for additive utilities \citep{Lee17,CaragiannisKuMo19,LiMaSc25}.
Moreover, the ECE algorithm offers considerable flexibility in the ordering of goods, the choice of unenvied agent to receive each good, and the order in which envy cycles are eliminated.\footnote{This flexibility has been utilized to obtain other properties in addition to EF1 \citep{BarmanKr20,MarkakisSa23,AmanatidisBiRe26,CelineSu26}.}

As discussed by \citet{LiptonMaMo04}, a standard implementation of the ECE algorithm runs in $O(n^3m)$ time, where $n$ and $m$ denote the number of agents and the number of goods, respectively.
In particular, allocating each good increases the number of envy edges by at most $n-1$, whereas eliminating each envy cycle decreases this number by at least one.
Since the envy graph is initially empty, at most $(n-1)m$ envy cycle eliminations occur.
Finding and eliminating an envy cycle takes $O(n^2)$ time, thereby yielding the $O(n^3m)$ running time.
\citet[Thm.~5.1]{OhPrSu21} observed that $O(nm)$ utility queries are sufficient for this implementation.
Subsequently, using techniques from dynamic weighted bipartite matching, \citet[Sec.~4]{MoritaShYa24} lowered the running time to $O(n^2m)$ for a version of ECE where bundles are reassigned through minimum-weight perfect matching computation rather than via explicit envy cycle elimination.

In this paper, we improve upon prior running times by showing that the ECE algorithm can be implemented in $O(nm)$ time.
Our implementation maintains a stack consisting of bundles based on the envy relations between their owners: the owner of each bundle (except the first) envies the owner of the preceding bundle on the stack.
If an envy cycle is encountered, the algorithm eliminates it by rotating the corresponding bundles among their owners and popping these bundles from the stack.
On the other hand, if an unenvied agent is found, the algorithm allocates the next good to her and resets the entire stack.
We remark that while the original implementation of \citet{LiptonMaMo04} eliminates all envy cycles before allocating the next good, this is unnecessary for the EF1 guarantee---as implicitly observed by \citet[Alg.~2]{AmanatidisAzBi23}, the algorithm can proceed as long as it identifies an unenvied agent.\footnote{If there is no unenvied agent, there must exist an envy cycle, so the algorithm can eliminate it first.}
Our implementation leverages this observation to achieve the $O(nm)$ running time. 
We also complement this result by establishing a matching lower bound: any deterministic algorithm that computes an ECE allocation requires $\Omega(nm)$ queries, and therefore $\Omega(nm)$ time, in the worst case.

\section{Preliminaries}
\label{sec:prelim}

For each positive integer~$k$, let $[k] \coloneqq \{1,\dots,k\}$.
Let $N = [n]$ be a set of $n\ge 2$ agents, and $M = \{g_1,\dots,g_m\}$ be a set of $m\ge 1$ goods.
A \emph{bundle} is a (possibly empty) subset of~$M$.
Each agent $i\in N$ has a utility function $u_i: 2^M\rightarrow \mathbb{R}_{\ge 0}$ such that $u_i(\emptyset) = 0$.
We assume that $u_i$ is \emph{monotone}, that is, $u_i(S)\le u_i(T)$ for any $S\subseteq T\subseteq M$; we do not require $u_i$ to be additive.
When analyzing the running time, we assume that an algorithm can query any agent's utility for any bundle in constant time.

An \emph{allocation} $A = (A_1,\dots,A_n)$ consists of $n$ bundles such that $A_i\cap A_j = \emptyset$ for any $i \ne j$; the allocation is \emph{complete} if $A_1\cup\dots\cup A_n = M$.
For any $i,j\in N$, agent~$i$ is said to envy agent~$j$ with respect to allocation~$A$ if $u_i(A_i) < u_i(A_j)$.
An allocation~$A$ satisfies \emph{envy-freeness up to one good (EF1)} if for any $i,j\in N$ with $A_j\neq \emptyset$, there exists $g\in A_j$ such that $u_i(A_i) \ge u_i(A_j\setminus\{g\})$.

The \emph{envy cycle elimination (ECE)} algorithm operates as follows.
Start with an empty allocation, and fix the ordering $g_1,\dots,g_m$ of the goods.
Maintain an \emph{envy graph}, which is a directed graph such that the set of vertices corresponds to the set of agents~$N$ and there is an edge from agent~$i$ to agent~$j$ exactly when $i$ envies~$j$ with respect to the current allocation.
An \emph{envy cycle} is a cycle in the envy graph.
At any point during the execution of the algorithm, two types of operations are allowed.
\begin{enumerate}
\item Allocate the next good to an unenvied agent.
\item Eliminate an envy cycle. Specifically, for an envy cycle $i_1\rightarrow i_2\rightarrow \dots\rightarrow i_k\rightarrow i_1$ with $k\ge 2$, we give agent $i_{\ell+1}$'s bundle to agent~$i_\ell$ for each $\ell\in[k-1]$, and give agent $i_1$'s bundle to agent~$i_k$.
\end{enumerate}
The algorithm can terminate when the allocation is complete.
Any (complete) allocation that can be returned by the ECE algorithm is called an \emph{ECE allocation}.
An ECE allocation always satisfies EF1 \citep{LiptonMaMo04}.
Note that if $m \le n$, any complete allocation that gives at most one good to each agent is an ECE allocation, since an agent with an empty bundle is always unenvied; hence, computing an ECE allocation is trivial in this case.
We will therefore assume from now on that $m > n \ge 2$.

\section{Algorithm}

In this section, we provide an $O(nm)$ implementation of the ECE algorithm.

\begin{theorem}
\label{thm:algorithm}
For monotone utilities, there exists a deterministic algorithm that always computes an ECE allocation in $O(nm)$ time.
\end{theorem}

\begin{proof}
We first describe the algorithm, then analyze its correctness and running time.

\medskip

\underline{Description}:
Create bundles $B_1,\dots,B_n$.
Initially, all of these bundles are empty, and each agent $i\in [n]$ holds bundle $B_i$.
We maintain arrays\footnote{Throughout the paper, we let the index of an array start from~$1$.} $\owner$, $\held$, and $\util$ of length~$n$ each.
For each $b\in[n]$, $\owner[b]$ indicates the agent who owns bundle~$B_b$.
For each $i\in [n]$, $\held[i]$ indicates the index of the bundle held by agent~$i$, and $\util[i]$ indicates agent~$i$'s utility for her current bundle, i.e., $u_i(B_{\held[i]})$.
Each bundle~$B_b$ is also associated with an array $\pairs[b]$ of capacity~$n$---which contains pairs $(i,w)$ indicating that agent~$i$ has utility~$w$ for~$B_b$---along with a pointer $\h[b]$.
Let $\size(\pairs[b])$ denote the number of filled entries in $\pairs[b]$.
The pointer $\h[b]$ always satisfies $1 \le \h[b] \le \size(\pairs[b]) + 1$.
Only the pairs at indices $\h[b], \h[b]+1, \dots, \size(\pairs[b])$ are said to be \emph{unconsumed}.
An unconsumed pair $(i,w)$ in $\pairs[b]$ is \emph{live} if $w > \util[i]$ (i.e., agent~$i$ envies the agent holding bundle~$B_b$), and \emph{stale} otherwise.

In addition, create a stack $P$ with capacity~$n$, which will hold bundle indices.
Let $\size(P)$ denote the number of entries in~$P$.
Maintain an array $\pos$ of length~$n$, where $\pos[b]$ indicates the position of index~$b$ in~$P$ (if $b$ is not in~$P$, $\pos[b]$ is set to $-1$).
When we push an index~$b$ into~$P$, $\pos[b]$ is set to $\size(P) + 1$, where we use the stack size before the push.
Conversely, when we pop $b$ from~$P$, $\pos[b]$ is reset to $-1$.

Let $r$ be an index that identifies the next good to be allocated.
Initially, set $r$ to~$1$ and $P$ to an empty stack. 
For each $i\in[n]$, set $\held[i]$ to $i$ and $\util[i]$ to $0$.
For each $b\in[n]$, set $\owner[b]$ to $b$, $B_b$ to an empty bundle, $\pairs[b]$ to an empty array, $h[b]$ to~$1$, and $\pos[b]$ to $-1$.
The algorithm then performs the following steps until it terminates with an allocation (in Step~3).
\begin{enumerate}
\item If the stack~$P$ is empty, push index~$1$ into~$P$.
Let $b$ be the index at the top of the stack.

\item While $\h[b] \le \size(\pairs[b])$, inspect the pair $(i,w) = \pairs[b][\h[b]]$.
If $w \le \util[i]$, the pair is stale (i.e., agent~$i$ no longer envies the owner of~$B_b$); increment the pointer $\h[b]$ by~$1$.
Repeat until either $\h[b] = \size(\pairs[b])+1$ (i.e., the array $\pairs[b]$ is exhausted) or $w > \util[i]$ for the current pair $(i,w)$ which is live.

\item If the array $\pairs[b]$ is exhausted, the owner $i^* = \owner[b]$ is unenvied.
In this case, we add the next good $g_r$ to $B_b$.
Query each agent~$i$'s utility $u_i(B_b)$, and set $\util[i^*] = u_{i^*}(B_b)$.
Empty $\pairs[b]$ and set $h[b]$ to~$1$.
For each $i\ne i^*$ such that $u_i(B_b) > \util[i]$, add a pair $(i,u_i(B_b))$ to $\pairs[b]$.
Pop the entire stack~$P$, resetting all $\pos$ entries to~$-1$.
Increment $r$ by $1$.

If all goods have been allocated, return the allocation~$A$ with $A_i = B_{\held[i]}$ for each $i\in [n]$.
Else, return to Step~1.

\item If the algorithm reaches this step, the array $\pairs[b]$ is not exhausted, and the current pair $(i,w) = \pairs[b][\h[b]]$ is live.
Let $b^* = \held[i]$ be the index of the bundle held by agent~$i$.\
\begin{enumerate}
\item If $b^*$ is not in the stack~$P$ (i.e., $\pos[b^*] = -1$), push $b^*$ into~$P$, update $\pos[b^*]$, and return to Step~1.
Note that the owner of $B_{b^*}$ (i.e., agent~$i$) envies the owner of $B_b$, so envy goes backward in the stack.
\item Suppose that $b^*$ is already in the stack~$P$ (i.e., $\pos[b^*] \ne -1$).
Let the stack contain entries $b_1,\dots,b_t$ in this order, and let $\ell = \pos[b^*]$ (i.e., $b_\ell = b^*$).
The suffix $b_\ell, b_{\ell+1},\dots,b_t$ corresponds to an envy cycle, which we will eliminate.
For each $c\in\{b_\ell,b_{\ell+1},\dots,b_t\}$, let $(i,w) = \pairs[c][\h[c]]$.
We give the bundle~$B_c$ to agent~$i$ by setting $\owner[c]$ to~$i$, $\held[i]$ to~$c$, and $\util[i]$ to~$w$, and increment $\h[c]$ by~$1$.
Pop $b_t, b_{t-1},\dots,b_\ell$ from~$P$, and reset the corresponding entries in $\pos$ to~$-1$.
Return to Step~1.
\end{enumerate}
\end{enumerate}

\medskip

\underline{Correctness}:
First, observe that for every pair $(i,w)$ in $\pairs[b]$ for some~$b$, it holds that $u_i(B_b) = w$.
Indeed, this is true when $(i,w)$ is added to $\pairs[b]$, and if a good is added to~$B_b$, the array $\pairs[b]$ is emptied and the pair $(i,w)$ thereby removed.
Also, whenever an agent~$i$ envies the owner of~$B_b$, there is an unconsumed pair $(i,w)$ in $\pairs[b]$.
This is because such a pair is added to $\pairs[b]$ upon the addition of the last good to~$B_b$, and the pair can only be consumed when either it is verified to be stale (i.e., agent~$i$ no longer envies the owner of~$B_b$) or agent~$i$ receives $B_b$ itself upon the elimination of an envy cycle (in which case the envy trivially disappears).
Note that agents' utilities never decrease during the execution of the algorithm, so a stale pair cannot become live later.
Therefore, if $\pairs[b]$ is exhausted (i.e., $\h[b] = \size(\pairs[b])+1$), this means that the owner of~$B_b$ is unenvied, and the next good can be added to~$B_b$.

For the stack~$P$, we maintain the invariants that all entries are distinct and the owner of each entry (except the first) envies the owner of the preceding entry.
In Step~4(a), the checks before the addition of~$b^*$ ensure distinctness as well as the envy relation.
In Step~4(b), distinctness is satisfied because we only pop entries, and the envy relations continue to hold for the remaining entries since the corresponding owners are unchanged.
Furthermore, when a new good is added, the entire stack is popped and the invariants trivially hold.
Hence, an envy cycle identified in Step~4(b) is valid, and the algorithm correctly eliminates this cycle.

\medskip

\underline{Running time}:
When a good is allocated, at most $n-1$ pairs are created, so the total number of pairs created during the execution of the algorithm is at most $(n-1)m = O(nm)$.
Each pair is consumed at most once.
Whenever an envy cycle of some length~$k$ is eliminated, $k$~pairs are consumed, so the total length of all eliminated envy cycles is $O(nm)$.
The time taken to eliminate an envy cycle is proportional to its length.
Entries in the stack~$P$ can also be popped due to a new good being allocated.
Since there are $m$ goods and at most $n$ entries in~$P$ are popped when a good is allocated, the number of entries popped in this way is also $O(nm)$.
The number of pushes is $O(nm)$ as well, since it can exceed the number of pops by at most~$n$.
The number of queries made is $n$ per good, that is, $nm$ in total.
Each iteration of the main loop of the algorithm ends with a good being allocated (Step~3), an entry being pushed into~$P$ (Step~4(a)), or an envy cycle being eliminated (Step~4(b)), so the number of iterations is $O(nm)$.
It follows that the entire algorithm runs in $O(nm)$ time.
\end{proof}

Since an ECE allocation always satisfies EF1, we obtain the following corollary.

\begin{corollary}
For monotone utilities, there exists a deterministic algorithm that always computes a complete EF1 allocation in $O(nm)$ time.
\end{corollary}

If the utilities are also \emph{additive}---that is, $u_i(S) = \sum_{g\in S}u_i(\{g\})$ for any $S\subseteq M$---a more restrictive query model is to only allow an algorithm to query an agent's utility for a single good \citep{LiMaSc25}.
We remark that the algorithm in \Cref{thm:algorithm} can be modified to yield an implementation with running time $O(nm)$ in this model as well.
In particular, we keep track of each agent~$i$'s utility for each bundle~$B_b$.
Whenever a good~$g$ is added to a bundle~$B_b$ (in Step~3), we query $u_i(\{g\})$ for all agents~$i$ and update $u_i(B_b)$ accordingly.
The rest of the algorithm proceeds in the same way as before.

\section{Lower Bound}

In this section, we establish the tightness of the $O(nm)$ bound for computing an ECE allocation.
We focus on deterministic algorithms but allow them to be adaptive, i.e., the next query of an algorithm can depend on the answers to its previous queries.

\begin{theorem}
\label{thm:lower-bound}
For monotone utilities, any deterministic algorithm that always computes an ECE allocation requires $\Omega(nm)$ queries, and therefore $\Omega(nm)$ time, in the worst case.
\end{theorem}

To prove \Cref{thm:lower-bound}, we will handle the cases $m \ge 2n$ and $n < m < 2n$ separately in the following two lemmas.
(Recall our assumption from \Cref{sec:prelim} that $m > n$.)

\begin{lemma}
\label{lem:large-m}
Let $m \ge 2n$.
For monotone utilities, any deterministic algorithm that always computes an ECE allocation requires at least $nm/12$ queries, and therefore $\Omega(nm)$ time, in the worst case.
\end{lemma}

\begin{proof}
Let $k = \lfloor m/n\rfloor \ge 2$, and consider any deterministic algorithm that always computes an ECE allocation.
Whenever the algorithm queries the utility of any agent~$i\in N$ for any set $S\subseteq M$, we answer that $u_i(S) = |S|$.
Observe that if the algorithm returns an allocation~$A$, every bundle in~$A$ must have size either $k$ or $k+1$.
Indeed, otherwise some agent~$j$ receives a bundle whose size is smaller than the bundle of another agent~$j'$ by at least~$2$.
If the utilities are such that $u_i(S) = |S|$ for all $i\in N$ and $S\subseteq M$, then agent~$j$ will envy agent~$j'$ by more than one good, contradicting the assumption that the algorithm always returns an ECE (and therefore EF1) allocation.

Fix $r\in [k-1]$ and an unordered pair of agents $i,j$, and let $A$ be the allocation that the algorithm returns.
We claim that the algorithm must make at least one of the following queries: $u_i(S)$ for some $S\subseteq A_j$ with $|S| = r$, or $u_j(S)$ for some $S\subseteq A_i$ with $|S| = r$.
Note that this claim implies the desired result, since the algorithm must then make at least $(k-1)\binom{n}{2}$ queries, and
\begin{align*}
(k-1)\binom{n}{2} = \left(\left\lfloor \frac{m}{n}\right\rfloor - 1\right)\binom{n}{2} \ge \frac{m}{3n}\cdot \frac{n(n-1)}{2} \ge \frac{m}{3n}\cdot \frac{n^2}{4} = \frac{nm}{12}.
\end{align*}
Here, the first inequality holds because $\lfloor x\rfloor - 1 \ge x/3$ for any $x \ge 2$, and the second inequality because $x-1 \ge x/2$ for any $x \ge 2$.

It remains to prove the claim.
Assume for contradiction that for some $r\in[k-1]$ and some pair of agents $i,j$, the algorithm does not make any query of the aforementioned type.
We complete the utility functions as follows.
The utility of agent~$i$ for a set $S\subseteq M$ is $|S|$, and it is incremented by $1/2$ if $|S| = r$ and $S\subseteq A_j$.
Similarly, the utility of agent~$j$ for a set $S\subseteq M$ is $|S|$, and it is incremented by $1/2$ if $|S| = r$ and $S\subseteq A_i$.
The utility of any other agent for a set $S\subseteq M$ is simply $|S|$.
Note that these utility functions are monotone.
Since the algorithm does not query $u_i(S)$ with $|S|=r$ and $S\subseteq A_j$, nor $u_j(S)$ with $|S|=r$ and $S\subseteq A_i$, the utility functions are also consistent with our answers.

We will show that the allocation~$A$ is not an ECE allocation, thereby obtaining the desired contradiction.
Suppose otherwise, that is, $A$ can be obtained by the ECE algorithm.
Observe that an agent~$a$ can envy another agent~$b$ only if $b$'s bundle is at least as large as $a$'s bundle, and if the two bundles have equal size, this size must be~$r$.
Hence, in any envy cycle, all bundles involved must be of size~$r$.
This means that once a bundle contains more than $r$ goods, its owner can no longer change.

Since the ECE algorithm only adds goods to bundles and permutes bundles among agents, we may track the bundles that end up as $A_i$ and $A_j$. 
Consider the first point at which either bundle grows from size~$r$ to $r+1$; this point must exist because both bundles eventually have size at least $k \ge r+1$.
Assume without loss of generality that at this point, $A_i$ grows from size~$r$ to $r+1$.
Since a bundle's owner cannot change after it has size at least $r+1$, the bundle~$A_i$ is already with agent~$i$ when it has size~$r$.
At this point, agent~$j$ cannot have more than $r$ goods, because if she does, then her bundle must be $A_j$ and grows from size~$r$ to $r+1$ before $A_i$ does.
Since agent~$j$'s bundle at this point cannot contain goods from (the final bundle) $A_i$, her utility for it is at most~$r$.
On the other hand, agent~$j$'s utility for agent~$i$'s bundle before its growth from size~$r$ to $r+1$ is $r + 1/2$.
Therefore, agent~$j$ envies agent~$i$, and so agent~$i$'s bundle should not grow from size~$r$ to $r+1$ at this point, a contradiction.
It follows that $A$ cannot be obtained by the ECE algorithm, completing the proof.
\end{proof}

\begin{lemma}
\label{lem:small-m}
Let $n < m < 2n$.
For monotone utilities, any deterministic algorithm that always computes an ECE allocation requires at least $nm/64$ queries, and therefore $\Omega(nm)$ time, in the worst case.
The same statement holds when ``ECE'' is replaced by ``complete EF1''.
\end{lemma}

\begin{proof}
Since an ECE allocation is complete and satisfies EF1, it suffices to prove the lemma for complete EF1.
Consider an algorithm that always computes a complete EF1 allocation.

Let $H\subseteq M$ be a set of $n+1$ goods, and call the goods in~$H$ \emph{active}.
All remaining goods are dummy goods which always yield zero marginal utility to every agent.
For each agent $i\in N$ and each good $g\in H$, let $x_{ig}\in\{1,1+\varepsilon\}$ for some $\varepsilon\in(0,1)$.
For any $i\in N$ and $S\subseteq M$, let $u_i(S) = x_{ig}$ if $S\cap H = \{g\}$ for some good~$g$, and $u_i(S) = |S\cap H|$ otherwise.
Note that the resulting utility functions are monotone regardless of the choices of $x_{ig}$.
We may assume that the algorithm knows the set of active goods~$H$ and the structure of the utility functions, but not the chosen values $x_{ig}$.
Hence, a query reveals additional information to the algorithm only if it involves a single active good; assume from now on that every query is of this form.

Observe that in a complete EF1 allocation, every agent must receive at least one active good.
Indeed, otherwise some agent receives no active good while another agent receives at least two active goods, so the former agent envies the latter agent even after removing one good.
Since there are $n+1$ active goods, one agent must receive two active goods while the remaining agents receive one active good each.

Let $r = n-1$ and $t = \lfloor r/4\rfloor$.
For each active good~$g$, we answer the first $t$ queries for distinct agents that reveal $x_{ig}$ with the value~$1$, and answer later queries that reveal $x_{ig}$ with the value $1+\varepsilon$.
Any repeated query always receives the same answer as before.
Let $h_1,h_2$ be the active goods in the unique bundle with two active goods, and let $i^*$ be the owner of this bundle.
By our answering procedure, among the $r$~agents who own a single active good, at most $t+t = 2t$ agents have a known utility of~$1$ for at least one of $h_1$ and $h_2$. 

Next, let $j$ be an agent who owns a single active good but has a known utility of~$1$ for neither $h_1$ nor~$h_2$, and let $g_j$ be the active good allocated to~$j$.
We claim that the algorithm must query $x_{jg_j}$ and receive the answer $1+\varepsilon$.
Indeed, otherwise we can choose the values $x_{ig}$ for all $i\in N$ and $g\in H$ so that they are consistent with our answers and $x_{jg_j} = 1$, $x_{jh_1} = 1+\varepsilon$, and $x_{jh_2} = 1+\varepsilon$; then, agent~$j$ envies agent~$i^*$ by more than one good.

The number of agents who own a single active good but have a known utility of~$1$ for neither $h_1$ nor~$h_2$ is at least $r - 2t \ge r/2$.
By our answering procedure, in order for such an agent to receive the answer $1+\varepsilon$ for her active good, the algorithm must query at least $t+1$ agents about her active good.
This means that the total number of queries is at least 
\begin{align*}
\frac{r}{2}\cdot (t+1)\ge \frac{r}{2}\cdot\frac{r}{4} = \frac{r^2}{8} = \frac{(n-1)^2}{8} \ge \frac{1}{8}\cdot\left(\frac{n}{2}\right)^2 = \frac{n^2}{32} > \frac{nm}{64},
\end{align*}
where the second inequality uses the assumption $n\ge 2$.
This completes the proof.
\end{proof}

\Cref{thm:lower-bound} is an immediate consequence of \Cref{lem:large-m,lem:small-m}.

\subsection*{Acknowledgments}

This work was partially supported by the Ministry of Education, Singapore, under the Academic Research Fund Tier 1 (FY 2026) grant number 251RES2604, and by an NUS Start-up Grant.

\subsection*{Declaration of Generative AI Use}

ChatGPT 6 Astra was used to help devise proofs with the author's guidance. 
The author wrote the entire article and takes full responsibility for its content.

\bibliographystyle{plainnat}
\bibliography{main}

\end{document}